\documentclass[a4paper,11pt]{article}

\usepackage{jheppub,amsmath,dsfont} 

\usepackage[T1]{fontenc} 

\newcommand{\black}{\color{black}}

\newcommand{\be}{\begin{equation}}
\newcommand{\ee}{\end{equation}}
\newcommand{\N}{\mathcal{N}}
\newcommand{\SO}{\text{SO}}

\title{\boldmath Gauge $\times$ Gauge  on Spheres}

\author[a]{L. Borsten,}
\author[a]{I. Jubb,}
\author[]{V. Makwana,}
\author[b,1]{S. Nagy\note{Corresponding author.}}

\affiliation[a]{School of Theoretical Physics, Dublin Institute for Advanced Studies,\\
10 Burlington Road, Dublin 4, Ireland}
\affiliation[b]{Centre for Astronomy \& Particle Theory,
University Park,
Nottingham,
NG7 2RD,
United Kingdom}

\emailAdd{leron@stp.dias.ie}
\emailAdd{ijubb@stp.dias.ie}
\emailAdd{visheshmakwana@gmail.com}
\emailAdd{silvia.nagy@nottingham.ac.uk}

\abstract{We introduce a convolution on a 2-sphere and use it to show that the linearised Becchi-Rouet-Stora-Tyutin  transformations and gauge fixing conditions of Einstein-Hilbert gravity coupled to a two-form and a scalar field, follow from the product  of two Yang-Mills theories. This provides an example of the convolutive product of gauge theories on a non-trivial background. By introducing  a time direction the product is shown to extend to the $D=1+2$ Einstein-static universe.}

\begin{document} 
\maketitle
\flushbottom

\section{Introduction}

Using a convolution on a 2-sphere we show that the linearised Becchi-Rouet-Stora-Tyutin (BRST) transformations and gauge fixing conditions of Einstein-Hilbert gravity coupled to a two-form and a scalar field\footnote{This is sometimes refered to as $\N=0$ supergravity. Specifically, it consists of a graviton coupled to a Kalb-Ramond (KR) 2-form and a dilaton as given by the low-energy effective field theory limit of the NS-NS (Neveu-Schwarz) sector of string theory.}, follow from the product  of two Yang-Mills theories. This provides an example of the convolutive product of gauge theories introduced in \cite{Anastasiou:2014qba} on a non-trivial background. 

The idea that spin-2 gravitons may be reformulated as the product, in some precise sense,  of  spin-1 gluons \cite{jordan1935neutrinotheorie, feynman2018feynman, Papini1965,Terazawa:1976eq}  has been re-invigorated in recent years,  the Weinberg-Witten theorem \cite{Weinberg:1980kq} notwithstanding.   This renaissance traces its origins  to the Kawai-Lewellen-Tye (KLT) scattering amplitude relations of  string theory  \cite{Kawai:1985xq}, but the crucial advance driving  recent  progress is the Bern-Carrasco-Johansson (BCJ) colour-kinematic duality and double-copy prescription \cite{Bern:2008qj, Bern:2010ue,  Bern:2010yg}. Given two gauge theories which satisfy  BCJ duality,  the   double-copy of their scattering amplitudes yields those of a gravitational theory to all orders in perturbation theory. BCJ duality for gluons has been established at tree-level from a number of perspectives \cite{Kiermaier:2010, BjerrumBohr:2010hn, Mafra:2011kj, Du:2016tbc} and has been generalised to include numerous (super) Yang-Mills theories  \cite{Bern:2009kd, Bern:2010ue, Bern:2010yg, Chiodaroli:2011pp, Bern:2012gh, Carrasco:2012ca, Damgaard:2012fb, Huang:2012wr, Bargheer:2012gv, Carrasco:2013ypa,  Chiodaroli:2013upa, Johansson:2014zca, Chiodaroli:2014xia, Chiodaroli:2015rdg, Chiodaroli:2015wal, Chiodaroli:2016jqw, Carrasco:2016ygv, Carrasco:2016ldy, Anastasiou:2016csv, Johansson:2017bfl, Johansson:2017srf, Azevedo:2017lkz, Anastasiou:2017nsz, Chiodaroli:2017ngp, Chiodaroli:2017ehv,  Chiodaroli:2018dbu,  Bern:2017ucb, Azevedo:2018dgo}, generating a wide variety of double-copy constructible  gravity theories (semi-classically, at least). Although BCJ duality remains conjectural at loop-level, there is a growing list of highly non-trivial examples \cite{Bern:2010ue,Bern:2010tq,Carrasco:2011mn,Bern:2011rj,BoucherVeronneau:2011qv,Bern:2012cd,Bern:2012gh, Oxburgh:2012zr,Bern:2012uf,Du:2012mt,Yuan:2012rg,Bern:2013uka, Boels:2013bi,Bern:2013yya,Bern:2013qca, Bern:2014lha, Bern:2014sna, Mafra:2015mja, Johansson:2017bfl,  Bern:2017ucb, Bern:2018jmv}.  This programme is certainly suggestive of a deep ``gravity $=$ gauge $\times$ gauge'' relation and, moreover,  has already dramatically advanced our understanding of perturbative  quantum gravity \cite{Bern:2009kd,Bern:2012cd, Bern:2012gh, Bern:2012uf, Bern:2013qca,Bern:2013uka, Bern:2014lha,Bern:2014sna, Bern:2015xsa, Bern:2018jmv}.  For a pedagogical introduction and  comprehensive review of these ideas and their applications see \cite{Carrasco:2015iwa, Bern:2019prr}. 

Given such successes it is natural to ask if the ``gravity $=$ gauge $\times$ gauge'' paradigm is strictly a property of scattering amplitudes alone.   There are a number of approaches one might take:   (i) the first thing one might consider is manifesting BCJ duality  at the level of the Lagrangian or field equations \cite{Bern:2008qj, Bern:2010ue, Monteiro:2011pc, BjerrumBohr:2012mg, Tolotti:2013caa, Monteiro:2013rya, Fu:2016plh, Cheung:2016say, Cheung:2016prv, Chen:2019ywi}; (ii) complementary to this is  the idea that  gravitational actions\footnote{Or, more likely, specifically those gravity theories that derive directly from the double-copy, in particular $\N=0$ supergravity.} can be recast into a form that, in some sense, factorises \cite{Bern:1999ji, Bern:2010yg, Hohm:2011dz, Cheung:2016say,  Cheung:2016prv, Cheung:2017kzx}; (iii) irrespective, one can apply the  BCJ double-copy paradigm to the construction of classical solutions in theories of gravity, such as black holes,  from gauge theory. This may take the guise of applying a classical  double-copy map to classical gauge theory solutions or extracting perturbative classical solutions from the double-copy of gauge theory  amplitudes  \cite{Monteiro:2014cda, Luna:2015paa,Ridgway:2015fdl,  Luna:2016due, White:2016jzc,Goldberger:2016iau,  Cardoso:2016ngt, Cardoso:2016amd, Luna:2016hge, Goldberger:2017frp, LopesCardoso:2018xes, Luna:2017dtq, Bahjat-Abbas:2017htu, Berman:2018hwd, Plefka:2018dpa, Bahjat-Abbas:2018vgo,Luna:2018dpt,Shen:2018ebu,  Cheung:2018wkq, Kosower:2018adc, CarrilloGonzalez:2019gof, Johansson:2019dnu,   Maybee:2019jus,Plefka:2019hmz,  Bern:2019nnu,  Bern:2019crd, Arkani-Hamed:2019ymq,Alawadhi:2019urr, Bah:2019sda, Plefka:2019wyg}; (iv) another approach is to seek a geometric and/or world-sheet understanding  of these relations through string theory \cite{BjerrumBohr:2009rd, Stieberger:2009hq, BjerrumBohr:2010hn,Schlotterer:2012ny, Ochirov:2013xba, Bjerrum-Bohr:2014qwa, He:2015wgf} or ambi-twistor strings and the scattering equations \cite{Cachazo:2013iea, Mason:2013sva, Adamo:2013tsa, Dolan:2013isa, Cachazo:2014xea,  Geyer:2015bja,  Geyer:2016wjx, Bjerrum-Bohr:2016axv,  Adamo:2017nia}.  

Here we consider a  further possibility:  a field theoretic  ``product'' of gauge theories \cite{Borsten:2013bp, Anastasiou:2014qba,Nagy:2014jza,Anastasiou:2013hba, Borsten:2015pla, Anastasiou:2015vba, Anastasiou:2016csv, Cardoso:2016ngt, Cardoso:2016amd, Anastasiou:2017nsz, Borsten:2017jpt, LopesCardoso:2018xes, Anastasiou:2018rdx}.  A covariant field theory product of arbitrary and independent gauge theories was introduced in \cite{Anastasiou:2014qba}.  For a Minkowski background, the local/global symmetries and equations of motion of the resulting gravity theory have been shown to follow from those of the gauge theory factors, to linear order, making crucial use of the the BSRT formalism  \cite{Anastasiou:2014qba, Anastasiou:2018rdx}, as reviewed in \autoref{rev}.  It can be used to construct, for example, supersymmetric (single/multi-centre) black hole solutions in $\N=2$ supergravity \cite{Cardoso:2016ngt, Cardoso:2016amd}, in the weak-field limit. The field theoretic product is \emph{a priori} independent from the BCJ double-copy, however it appears to be  consistent with it in the sense that the double-copy amplitudes correspond to the theory obtained from the field product \cite{Anastasiou:2017nsz}. This not merely a statement that the spectra match; it requires that the symmetries and couplings agree and for $\N<4$ this did not have to be the case. The product also makes use of a bi-adjoint (and bi-fundamental) scalar field closely related, as the convolutive pseudo-inverse, to the $\phi^3$-theory appearing in the double-copy in various forms \cite{Hodges:2011wm, Vaman:2010ez, Cachazo:2013iea, Monteiro:2013rya, Cachazo:2014xea,Monteiro:2014cda, Chiodaroli:2014xia,  Naculich:2014naa, Luna:2015paa, Naculich:2015zha, Chiodaroli:2015rdg, Luna:2016due, White:2016jzc,  Cheung:2016prv, Chiodaroli:2017ngp, Brown:2018wss}. Remarkably, the product is non-trivial in contexts where it has no \emph{a priori} right  to be applied, such as in the absence of a perturbative limit \cite{Borsten:2017jpt, Borsten:2018jjm}. This has been used to clarify aspects of known theories \cite{Borsten:2017jpt} as well as to discover previously unknown theories \cite{Borsten:2018jjm}, building on \cite{Ferrara:2018iko}.  To include higher-order interactions the field product can be used in conjunction with the perturbative classical double-copy of \cite{Luna:2016hge}, generalised to accommodate the BRST formalism  \cite{BorstenXXX}. 

A natural generalisation in any of the above contexts is to curved background spacetimes. There are two obvious perspectives on this. The first is to consider a non-trivial gauge field background on a \emph{flat} spacetime. The idea is that the gauge background generates a non-trivial spacetime background in the gravity theory;  gluon scattering on a non-trivial gauge background  generates graviton scattering on a non-trivial spacetime background. This has been successfully explored  in the context of the ambi-twistor string formalism for plane-wave gauge backgrounds \cite{Adamo:2017nia}. More generally, a key question here is how to correctly identify the non-perturbative  spacetime background from the non-perturbative gauge theory background(s). When the desired spacetime background is Kerr-Schild one might be able to use the classical  double-copy of \cite{Monteiro:2014cda} to identify the corresponding gauge background and then construct the associated  amplitudes through a generalisation of BCJ duality and the double-copy. This is clearly a challenging proposition and the general principles remain to be established.  Alternatively, one could consider gauge theory amplitudes on a curved spacetime background in the first place. In the strongest form of the ``gravity $=$ gauge $\times$ gauge'' proposal, one might hope to derive spacetime geometry without putting it into the gauge theory factors at the outset, although what perspective will ultimately emerge remains an open question. Irrespective, it is a perfectly sensible and interesting question to consider what would one get if one were able to apply BCJ duality and the double-copy, or any other incarnation of ``gravity $=$ gauge $\times$ gauge'', on curved backgrounds.  It is not hard to envisage various possible applications and lessons, for example graviton scattering on (amenable) curved backgrounds. 

It is this second path we consider here in the context of the field theory product. Specifically, we introduce a convolution product defined on a 2-sphere. It is required to be covariant with respect to the isometries of the sphere, just as  the   product in Minkowski spacetime  is for the   Poncar\'e group. Applying the product to the BRST  formulation of two linearised pure Yang-Mills theories we obtain a graviton, KR 2-form and dilaton on the sphere. It is shown that the linearised BRST transformations and gauge-fixing conditions of the gravitational fields follow directly from the gauge theory factors. We begin in \autoref{rev} with a review of the key ingredients of the BRST field theory product in flat spacetime. We then introduce the convolution on the sphere in \autoref{sphere}. The key technical ingredient is the introduction of a convolution for tensor fields that is multiplicative in Fourier space. This formalism is then applied to two gauge theories on the sphere, which are shown to generate the local symmetries (BRST transformations) of Einstein--KR 2-form--dilaton gravity in \autoref{dcsphere}. Moreover, the gauge choice in the factors determines the gauge choice in the product, in  a precise sense. Finally, in \autoref{static} we introduce time and consider as the simplest example the $D=3$ spacetime dimensional Einstein static universe. To conclude we consider the possible generalisations, in particular the extension to all group manifolds taking $S^3$ as the simplest example. 

\section{Review of BRST field product in a flat background}\label{rev}

Following \cite{Anastasiou:2014qba, Anastasiou:2018rdx} we consider here the field theory product defined by
\be\label{def}
f\circ  \tilde{f}:=\langle\langle f , \Phi , { \tilde{f}}\rangle\rangle.
\ee
Here, $f, \tilde{f}$ are arbitrary spacetime fields valued in $\mathfrak{g}$ and $\tilde{\mathfrak{g}}$, respectively, which are the Lie algebras corresponding to the gauge groups $G$ and $\tilde{G}$. The ``spectator'' field $\Phi=\Phi^{a\tilde{a}}T_a\otimes \tilde{T}_{\tilde{a}}$ is a  $G \times \tilde{G}$ bi-adjoint valued scalar.   
 Here   $\langle\langle~,~,~\rangle\rangle$ is a trilinear trace form constructed from the negative-definite trace forms of $\mathfrak{g}, \tilde{\mathfrak{g}}$, which in the standard  basis is simply,
 \be
 \langle\langle X , Y , \tilde{X}\rangle\rangle = X_a\cdot Y^{a\tilde{a}}\cdot \tilde{X}_{\tilde{a}}\ , 
 \ee
where the $\cdot$  product denotes an associative  convolutive inner tensor product with respect to the Poincar\'e group
\be
\label{flat_convo_def}
[f\cdot g](x)=\int d^Dy f(y) \otimes g(x-y).
\ee
The convolution reflects the fact that the amplitude relations are multiplicative in momentum space. For sufficiently well-behaved functions the convolution  obeys,
 \be
\partial_{\mu}[f\cdot g](x)=[\partial_{\mu}f\cdot g](x)=[f\cdot \partial_{\mu}g](x).
\ee
  The double trace form accounts for the gauge groups, while the spectator field allows for arbitrary and independent   $G$ and $\tilde{G}$. Of course, it is closely related to the bi-adjoint scalar $\phi$ of the BCJ zeroth-copy \cite{Hodges:2011wm, Vaman:2010ez, Cachazo:2013iea,  Monteiro:2013rya}. It can be considered as its convolutive pseudo-inverse $\Phi = \phi^{-1}$, where $\phi\cdot\Phi\cdot\phi=\phi$. This implies that $\Phi$ has mass-dimension $(3D+2)/2$, which also ensures the mass-dimensions of the product fields are consistent. It also implies that it transforms inversely  under translations. This precisely compensates for the fact that under $x\mapsto x+a$, $[f\cdot g](x)\mapsto [f\cdot g](x+2a)$, i.e.~$[f\circ  {g}](x)\mapsto [f\circ  {g}](x+a)$. Note the circle product can be generalised to include fundamental matter fields, by including a bi-fundamental scalar field \cite{Anastasiou:2016csv}.

Having introduced the covariant product, let us consider the case of two pure Yang-Mills theories. The field-theoretic product of two gauge potentials, $A_\mu$ and  $\tilde{A}_\nu$,   is given by
\begin{equation}\label{product}
[A_\mu \circ   \tilde{A}_\nu](x) = g^2 [A_\mu^a \cdot \Phi_{a\tilde{a}} \cdot \tilde{A}_\nu^{\tilde{a}}](x).
\end{equation}
Naive on-shell counting shows that the product of two  Yang-Mills  gauge fields $A_\mu^a$ and $\tilde{A}_\nu^{\tilde{a}}$ yields a graviton $h_{\mu\nu}$, the Kalb-Ramond (KR) two-form $B_{\mu\nu}$ and a dilaton $\varphi$. Tree-level amplitudes fix the action to be that of $\N=0$ supergravity.  However, a number of significant issues have been identified in relation to this construction in the context of off-shell or classical approaches:
\begin{itemize}
\item It is generally difficult to disentangle the graviton and dilaton degrees of freedom \cite{LopesCardoso:2018xes,Luna:2017dtq}. A simple way to  see this is presented in \cite{LopesCardoso:2018xes}. Let  $j_\mu$ and $\tilde{j}_\mu$ be the sources of the   Yang-Mills  e.o.m., $j_{\mu\nu}^{(h)}$ the graviton source and $j^{(\varphi)}$ the dilaton source. Then we have 
\be 
\label{source_issue}
j^{(\varphi)}\propto j_{\ \rho}^{(h) \rho}\propto \tfrac{1}{\square}j_\rho\circ \tilde{j}^\rho \ .
\ee
Thus we see that the graviton and dilaton sources are not independent. We can interpret this as a constraint on gravitational theories that  admit a double copy description, appearing already at the linear order. This is a general feature of the classical BCJ double-copy, and not a consequence of the set-up  in  \cite{LopesCardoso:2018xes}.  
\item There is a mismatch between the on-shell and off-shell d.o.f.~mapping. Generally, an off-shell  $A_\mu\times\tilde{A}_\nu$ product does not carry a sufficient number of degrees of freedom to describe the graviton--two-form--dilaton system off-shell \cite{LopesCardoso:2018xes,Anastasiou:2018rdx}. The issue persists with the addition of supersymmetry \cite{Anastasiou:2014qba,Siegel:1988qu,Siegel:1995px}.   

\item It is not clear how the required gauge-for-gauge freedom of the KR 2-form is to be  accommodated \cite{BorstenXXX}. 

\item The  classical double-copy is usually formulated with some specific gauge fixing on both the  Yang-Mills  and the gravity side. However, there is no general procedure determining a mapping between these corresponding gauge choices - this can lead to issues, particularly when studying off-shell or gauge-dependent objects \cite{Plefka:2019hmz}.
\end{itemize} 
The BRST dictionary in \cite{Anastasiou:2018rdx}  resolves the above issues by taking products of sets of fields $(A_\mu,c^\alpha)$ and $(\tilde{A}_\mu,\tilde{c}^\alpha)$. Here $c^1=c$ and $c^2=\bar{c}$ are the Fadeev-Popov ghost and antighost, respectively. The off-shell d.o.f.~of the  $(A_\mu,c^\alpha)\times(\tilde{A}_\mu,\tilde{c}^\alpha)$ product can now be seen to correspond to those of the linearised BRST systems for the graviton, two-form and dilaton\footnote{Note that the d.o.f. counting is now graded by ghost number - see  \cite{Anastasiou:2018rdx,BorstenXXX} for details.}. It also naturally incorporates the ghost and ghost-for-ghost transformations \cite{Anastasiou:2018rdx,BorstenXXX}, addressing the penultimate concern above. 

We will describe below how the BRST procedure resolves the source issue \eqref{source_issue}, at the same time as giving a gauge mapping algorithm between between pure Yang-Mills  theory and gravity coupled to a KR 2-form and a dilaton.

The general form of the BRST action for a field $f$, with an irreducible gauge symmetry is schematically
\be 
S_{\text{BRST}}=\int d^D x\left(\mathcal{L}_0[f]+b\left(G[f]-\tfrac{\xi}{2}b\right)-\bar{c}Q\left(G[f]\right) \right)-f j^{(f)}+\bar{j}c+\bar{c}j \ ,
\ee
where $\mathcal{L}_0[f]$ is the classical action for the field $f$, $G[f]$ is the gauge-fixing functional and $b$ is the Lautrup-Nakanishi Lagrange multiplier field. Finally, $Q$ is a homological vector field and functions on this enlarged space of fields, $(A, c, \bar{c}, b)$, form a chain complex; its $Q$-cohomology characterises the physical observables. For reducible gauge symmetries there will be additional ghost-for-ghost terms. For a review of the BRST procedure, see \cite{Kugo:1979gm,Henneaux:1992ig, Gomis:1994he,Zoccali:2018pty}.

Note that, unlike in the standard treatment, we have coupled sources to both the physical field and the ghosts.
Specialising to linearised  Yang-Mills  fields and their ghosts, the e.o.m.~arising from the above are
\be 
\begin{aligned}
\partial^\mu F_{\mu\nu}+\tfrac{1}{\xi}\int d^D y G[A]\frac{\delta G[A]}{\delta A^\nu}=&j_\nu, \\
Q\left(G[A]\right)=&j,\\
\int d^Dy\bar{c}\frac{\delta Q\left(G[A]\right)}{\delta c}=&\bar{j} \ ,
\end{aligned}
\ee
whereas for the gravity fields in the linearised approximation we have
\be 
\label{eom_gravity_flat}
\begin{aligned}
R_{\mu\nu}+\left(\delta_\mu^\alpha\delta_\nu^\beta-\tfrac{1}{D-2}\eta_{\mu\nu}\eta^{\alpha\beta} \right)
\tfrac{1}{\xi^{(h)}}\int d^D y G^\rho[h,\varphi]\frac{\delta G_\rho[h,\varphi]}{\delta h^{\alpha\beta}}=&j_{\mu\nu}^{(h)},\\
\partial^\rho H_{\rho\mu\nu}[B]+\tfrac{1}{\xi^{(B)}}\int d^D y G^\rho[B,\eta]\frac{\delta G_\rho[B,\eta]}{\delta B^{\mu\nu}}=&j_{\mu\nu}^{(B)},\\
\square\varphi=&j^{(\varphi)}, 
\end{aligned}
\ee
where the gauge fixing functional for the graviton is allowed to depend on the dilaton, and the gauge fixing functional of the two-form contains the second-level ghost $\eta$ of ghost number zero \cite{Henneaux:1992ig, Gomis:1994he, Zoccali:2018pty}. 

Not allowing for non-local derivative operators\footnote{Of course, can can relax this condition, and consider alternative  restrictions by making various assumptions about the ans\"atze for the field and source dictionaries. See \cite{Anastasiou:2018rdx}.}, the field theory dictionary is  uniquely fixed by the BRST transformations at linear order in Einstein frame. A basic requirement is that the BRST symmetries of  Yang-Mills  system,
\be 
Q A_\mu=\partial_\mu c,\quad Qc=0,\quad Q\bar{c}=\tfrac{1}{\xi}G(A),
\ee  
induce the correct symmetries for the gravitational fields, those  relevant to the present discussion being:
\be 
\label{flat_grav_BRST}
\begin{array}{lllllllll}
Qh_{\mu\nu}&=&2\partial_{(\mu}c_{\nu)},\quad& Qc_\mu&=&0, \quad& Q\bar{c}_\mu&=&\tfrac{1}{\xi^{(h)}}G_\mu[h,\varphi],  \\[5pt]
QB_{\mu\nu}&=&2\partial_{[\mu}d_{\nu]},\quad& Qd_\mu&=&\partial_\mu d, \quad&  Q\bar{d}_\mu&=&\tfrac{1}{\xi^{(B)}}G_\mu[B,\eta], \\[5pt]
Q\varphi&=&0.\\
\end{array}
\ee
Let us now make a choice of gauge fixing functional on the  Yang-Mills  side, and set
\be
\label{flat_gauge_choice}
G[A]\equiv\partial^\mu A_\mu,\quad G[\tilde{A}]\equiv\partial^\mu \tilde{A}_\mu .
\ee 
The gauge fixing functionals for $h_{\mu\nu}$ and $B_{\mu\nu}$ are to be determined through our procedure, as described below. The most general dictionary in the absence of $\Box^{-1}$ terms and  compatible with symmetries is:
\be
\label{flat_simple_dict}
\begin{aligned}
h_{\mu\nu}=&A_\mu\circ\tilde{A}_\nu+A_\nu\circ\tilde{A}_\mu+a\eta_{\mu\nu}\left(A^\rho\circ\tilde{A}_\rho + \xi c^\alpha\circ\tilde{c}_\alpha\right),\\
B_{\mu\nu}=&A_\mu\circ\tilde{A}_\nu-A_\nu\circ\tilde{A}_\mu, \\
\varphi=&A^\rho\circ\tilde{A}_\rho + \xi c^\alpha\circ\tilde{c}_\alpha,
\end{aligned} 
\ee
where we have introduced the OSp(2) ghost singlet 
\be 
\label{ghost_singlet}
c^\alpha\circ\tilde{c}_\alpha=c\circ\tilde{\bar{c}}-\bar{c}\circ\tilde{c}.
\ee
We can immediately read off the graviton and two-form ghost dictionaries,
\be
\begin{aligned}
c_\mu=&c\circ\tilde{A_\mu}+A_\mu\circ\tilde{c},\\
d_\mu=&c\circ\tilde{A_\mu}-A_\mu\circ\tilde{c},
\end{aligned} 
\ee
from which the antighost dictionaries follow:
\be
\begin{aligned}
\bar{c}_\mu=&\bar{c}\circ\tilde{A_\mu}+A_\mu\circ\tilde{\bar{c}},\\
\bar{d}_\mu=&\bar{c}\circ\tilde{A_\mu}-A_\mu\circ\tilde{\bar{c}}.
\end{aligned} 
\ee
The BRST transformations of the above are
\be \label{Qag}
\begin{aligned}
Q\bar{c}_\mu=&\tfrac{1}{\xi}\left[\partial^\rho A_\rho\circ\tilde{A}_\mu +A_\mu\circ\partial^\rho\tilde{A}_\rho \right]
+\partial_\mu c^\alpha\circ\tilde{c}_\alpha,\\
Q\bar{d}_\mu=&\tfrac{1}{\xi}\left[\partial^\rho A_\rho\circ\tilde{A}_\mu -A_\mu\circ\partial^\rho\tilde{A}_\rho \right]
-\partial_\mu\left[c\circ\tilde{\bar{c}}+\bar{c}\circ\tilde{\bar{c}} \right].
\end{aligned} 
\ee
From $Q\bar{d}_\mu$ we learn that the ghost-number-zero ghost-for-ghost field $\eta$ required for the KR gauge-fixing should be identified with $\xi\left(c\circ\tilde{\bar{c}}+\bar{c}\circ\tilde{\bar{c}}\right)$. This is the ghost-number-zero component of the triplet $c^{(\alpha}\circ\tilde{{c}}^{\beta)}$. The ghost number +2 and -2 components are the ghost-for-ghost field $d$ and its antighost, respectively. Then \eqref{Qag} allows us to read off the gauge-fixing functionals for the graviton and two-form, by making use of \eqref{flat_grav_BRST} and inverting the dictionaries \eqref{flat_simple_dict},
\be 
\label{flat_grav_gauge_simple}
\begin{aligned}
G_\mu[h,\varphi]=&\partial^\nu h_{\nu\mu}-\tfrac{1}{2}\partial_\mu h +\left(1+\tfrac{D-2}{2}a\right)\partial_\mu \varphi,\\
G_\mu[B,\eta]=&\partial^\nu B_{\nu\mu}-\partial_\mu\eta,
\end{aligned}
\ee
where we imposed $\xi=\xi^{(h)}=\xi^{(B)}$ to ensure  the $\xi$-independence of the gauge-fixing functionals. Demanding that the diffeomorphism gauge-fixing is independent of the dilaton fixes $a=2/(D-2)$ so that the  dictionary (in the absence of $\Box^{-1}$ terms) is determined uniquely. 

Finally, making use of the e.o.m. \eqref{eom_gravity_flat}, we can write source dictionaries:
\be 
\begin{aligned}
j_{\mu\nu}^{(h)}=&2 \tfrac{1}{\square}\black j_{(\mu}\circ\tilde{j}_{\nu)}+\left[\tfrac{2\xi +3}{\xi}+\tfrac{(D-2)a(2\xi+5)}{2\xi} \right] \tfrac{\partial_\mu\partial_\nu}{\square^2} j^\rho\circ\tilde{j}_\rho\\
&+\left[-1+\tfrac{(D-2)a(2\xi+5)}{2} \right] \tfrac{\partial_\mu\partial_\nu}{\square^2} j^\alpha\circ\tilde{j}_\alpha \\
&+ \left[-\tfrac{(\xi+1)(5\xi+3)}{\xi}+\tfrac{(D-2)a(\xi^2-1)(2\xi+5)}{2\xi} \right]\tfrac{\partial_\mu\partial_\nu}{\square^3}\partial^\rho j_\rho\circ\partial^\sigma \tilde{j}_\sigma\\
&+
2\tfrac{1}{\square^2}\left[\partial^\rho j_\rho\circ\partial_{(\mu}\tilde{j}_{\nu)}+\partial_{(\mu}j_{\nu)}\circ\partial^\rho \tilde{j}_\rho\right]\\
&+a\tfrac{1}{\square}\eta_{\mu\nu}j^\rho\circ\tilde{j}_\rho+a\xi\tfrac{1}{\square}\eta_{\mu\nu}j^\alpha\circ\tilde{j}_\alpha+ a(\xi^2-1)\eta_{\mu\nu}\tfrac{1}{\square^2} \partial^\rho j_\rho\circ\partial^\sigma \tilde{j}_\sigma,\\
j_{\mu\nu}^{(B)}=&2\tfrac{1}{\square}j_{[\mu}\circ \tilde{j}_{\nu]}+2\tfrac{1}{\square^2}\left[\partial_{[\nu}j_{\mu]}\circ\partial^\rho\tilde{j}_\rho-\partial^\rho j_\rho\circ\partial_{[\nu}\tilde{j}_{\mu]} \right],\\
j^{(\varphi)}=&\tfrac{1}{\square}j^\rho\circ\tilde{j}_\rho + (\xi^2-1)\tfrac{1}{\square^3}\partial^\rho j_\rho\circ\partial^\sigma\tilde{j}_\sigma+\xi\tfrac{1}{\square}j^\alpha\circ\tilde{j}_\alpha.
\end{aligned}
\ee
Note the source issue pointed out in \eqref{source_issue} is now resolved, as one can choose ghosts such that the dilaton is set to vanish, without affecting the dynamics of the graviton.

This procedure can be extended to higher orders by exploiting the connection between the perturbative expansion of the equations of motion and scattering amplitudes, in conjunction with the BCJ construction \cite{BorstenXXX}.

\section{Convolutions on $S^2$}\label{sphere}
\subsection{Scalar Convolution}
The convolution of scalar functions on $S^2$ is described in \cite{Driscoll:1994cca}. We work with the following conventions for the spherical Fourier transform of a function $f(\theta,\phi)$ and  its inverse:
\be
\begin{aligned}
f_l^m=&\int_{S^2}f(\theta,\phi)\bar{Y}_l^m (\theta,\phi)\sin\theta d\theta d\phi,\\
f(\theta,\phi)=&\sum_{l\in\mathbb{N}, |m|\leq l} f_l^mY_l^m (\theta,\phi).
\end{aligned} 
\ee 
Note, the spherical harmonics $Y_l^m$ can be  defined through
\be 
\label{def_Y}
Y_l^m(\theta,\phi)=\sqrt{\tfrac{(2l+1)}{4\pi}}\bar{D}^l_{m0}(\phi,\theta,\psi),
\ee
where $D^l_{mn}$ are the usual Wigner matrices. Here $\psi$ is arbitrary because we are selecting the 0'th column of the matrix; in particular we could set $\psi=0$. 

In \cite{Driscoll:1994cca} the left-convolution was defined as:
\be
\label{DH_convo}
\begin{aligned}
\,[k\cdot f] (\theta,\phi)=&\left(\int_{g\in\text{SO(3)}}dg~ k(g\eta)\Lambda(g)\right)f(\theta,\phi)\\
=&\int_{g\in\text{SO(3)}}dg~ k(g\eta) f\left(g^{-1}(\theta,\phi)\right)dg,
\end{aligned} 
\ee
where $\eta=(0,\cdot)$ is identified as the North Pole and $\Lambda(g)$ is an operator induced by the action of SO(3) on the sphere. Note that \eqref{DH_convo} has the same structure as the flat-space convolution \eqref{flat_convo_def}  - in both cases we integrate over the group acting transitively on the respective marked manifold: translations for flat space and SO(3) for the sphere.

A fundamental property required of convolutions is factorisation in the dual space. To check this, we first rewrite the convolution by employing the Euler angle description of the SO(3) action on spheres,
\be
\begin{aligned}
\,[k\cdot f](\theta,\phi)=&\int k\left(R_{(\alpha,\beta,\gamma)}\eta\right)f\left(R_{(-\gamma,-\beta,-\alpha)}(\theta,\phi)\right)\sin\beta d\alpha d\beta d\gamma\\
=&\int k\left(\beta,\alpha\right)f\left(R_{(-\gamma,-\beta,-\alpha)}(\theta,\phi)\right)\sin\beta d\alpha d\beta d\gamma.
\end{aligned}
\ee
Then we compute:
\be
\label{fact_DH_1}
\begin{aligned}
\left(k\cdot f \right)_l^m=&\int k\cdot f(\theta,\phi)\bar{Y}_l^m (\theta,\phi)\text{sin}\theta d\theta d\phi\\
=&\sum_{\substack{l'\in\mathbb{N}, |m'|\leq l'\\ l''\in\mathbb{N}, |m''|\leq l''}}k^{m'}_{l'}f_{l''}^{m''}\int
Y_{l'}^{m'}(\beta,\alpha)Y_{l''}^{m''}\left(R_{(-\gamma,-\beta,-\alpha)}(\theta,\phi)\right)\bar{Y}_l^m(\theta,\phi)dv,
\end{aligned} 
\ee 
where $dv=\sin\beta \sin\theta d\alpha d\beta d \gamma d\theta d\phi$. 
Then, using \eqref{def_Y} and the fact that the $D$'s are representation matrices of SO(3), we have:
\be 
\begin{aligned}
Y_{l''}^{m''}\left(R_{(-\gamma,-\beta,-\alpha)}(\theta,\phi)\right)=&
\sqrt{\tfrac{(2l+1)}{4\pi}}\bar{D}^{l''}_{m''0}\left(R_{(-\gamma,-\beta,-\alpha)}(\theta,\phi)\right)\\
=&\sqrt{\tfrac{(2l+1)}{4\pi}}\sum_{\tilde{m}}\bar{D}_{m''\tilde{m}}^{l''}(-\gamma,-\beta,-\alpha)\bar{D}_{\tilde{m}0}^{l''}(\phi,\theta,0)\\
=&\sum_{\tilde{m}}D_{\tilde{m}m''}^{l''}(\alpha,\beta,\gamma)Y_{l''}^{\tilde{m}}(\theta,\phi).
\end{aligned}
\ee
Plugging the above back into \eqref{fact_DH_1} and performing the integral over $\theta,\phi$, we get
\be 
\left(k\cdot f \right)_l^m=
\sum_{\substack{l'\in\mathbb{N}, |m'|\leq l'\\  |m''|\leq l}}k^{m'}_{l'}f_{l}^{m''}
\int Y_{l'}^{m'}(\beta,\alpha) D^l_{mm''}(\alpha,\beta,\gamma)\sin\beta d\alpha d\beta d\gamma,
\ee
using the orthogonality relations
\be\label{ortho_Y}
  \int_{S^2}Y_l^m (\theta,\phi) \bar{Y}_{l'}^{m'} (\theta,\phi)=\delta_{ll'}\delta_{mm'}. 
\ee

The integral over $\gamma$ gives $2\pi\delta_{m''0}$ so finally we have
\be 
\begin{aligned}
\left(k\cdot f \right)_l^m=&2\pi\sum_{l'\in\mathbb{N}, |m'|\leq l'} k^{m'}_{l'}f_{l}^{0}
\int Y_{l'}^{m'}(\beta,\alpha) D^l_{m0}(\alpha,\beta,0)\sin\beta d\alpha d\beta \\
=&2\pi\sqrt{\tfrac{4\pi}{2l+1}}\sum_{l'\in\mathbb{N}, |m'|\leq l'} k^{m'}_{l'}f_{l}^{0}
\int Y_{l'}^{m'}(\beta,\alpha) \bar{Y}_l^m(\beta,\alpha)\sin\beta d\alpha d\beta \\
=&2\pi\sqrt{\tfrac{4\pi}{2l+1}} k^{m}_{l}f_{l}^{0},
\end{aligned}
\ee
where to get to the second and third lines we used \eqref{def_Y}. Thus we have verified the factorisation property of the convolution integral. Note that the factorisation in Fourier space is not symmetric. This is a reflection of the fact that the have defined a left-convolution. One could also define a distinct right-convolution product. The situation is different from flat space, where the two definitions will be equivalent, due to the commutativity of translations.   

\subsection{Tensor Convolution}
We would now like to extend the spherical convolution \eqref{DH_convo} to tensor fields. We find it useful to first recast tensors on $S^2$ as scalars on SO(3), as described in \cite{Gelfand1963,Burridge1969}. Let $M_{\alpha_1...\alpha_p}(\mathbf{x})$ be a tensor field in $\mathds{R}^3$ and $g\in \text{SO(3)}$. We define 
\be 
\label{Burridge_basis}
\mathcal{M}_{\alpha_1...\alpha_p}(r,g)=g_{\alpha_1\beta_1}...g_{\alpha_p\beta_p}
M_{\beta_1...\beta_p}\left(rg^{-1}\mathbf{e_3} \right).
\ee
Geometrically, we can interpret the above in the following way. Consider an orthonormal frame $(\mathbf{i_1},\mathbf{i_2},\mathbf{i_3})$ parallel to the standard Cartesian frame, but with origin at the North Pole $\eta$. Let  $(\mathbf{j_1},\mathbf{j_2},\mathbf{j_3})$ be the frame obtained by rotating $(\mathbf{i_1},\mathbf{i_2},\mathbf{i_3})$ with $g^{-1}$. Then $\mathcal{M}_{\alpha_1...\alpha_p}$ are the components of $M$ at the point $g^{-1}\eta$, in the frame $(\mathbf{j_1},\mathbf{j_2},\mathbf{j_3})$. It is straightforward to check that each component $\mathcal{M}_{\alpha_1...\alpha_p}$ transforms as a scalar under rotations:
\be 
\begin{aligned}
\mathcal{M}'_{\alpha_1...\alpha_p}(r,g)=&g_{\alpha_1\beta_1}g'_{\beta_1\gamma_1}...g_{\alpha_p\beta_p}g'_{\beta_p\gamma_p}\ M_{\gamma_1...\gamma_p}\left(r\left(g'\right)^{-1}g^{-1}\mathbf{e_3} \right)\\
=&\mathcal{M}_{\alpha_1...\alpha_p}(r,gg').
\end{aligned}
\ee     
We also find it useful to perform a change of basis
\be 
\label{go_to_helicity_basis}
\mathcal{M}_{a_1...a_p}(r,g)=C_{a_1 \alpha_1}...C_{a_p \alpha_p} \mathcal{M}^{\alpha_1...\alpha_p}(r,g),  
\ee
where
\be 
C_{\alpha a}=
\begin{pmatrix}
\tfrac{1}{\sqrt{2}} &0  &-\tfrac{1}{\sqrt{2}}  \\ 
\tfrac{i}{\sqrt{2}}  &0  &\tfrac{i}{\sqrt{2}}  \\ 
0 & 1 & 0
\end{pmatrix}.
\ee
This can be thought of as going to a helicity basis; here $\alpha$ runs over $-1,0,+1$. The components of $\mathcal{M}^{a_1...a_p}(r,g)$ are similarly obtained by acting with the conjugate $\bar{C}_{\alpha a}$. Then, noting that
\be 
T^1_{ab}=\bar{C}_{\alpha a}\  g_{\alpha\beta}\ C_{\beta_b},
\ee
where $T$ is related to the usual Wigner matrices via
\be 
\label{gen_sph_harm}
T^l_{ab}(g)=(-1)^{b-a}\bar{D}^l_{ba}(g),
\ee
we finally find
\be 
\label{sum_expl}
\mathcal{M}_{a_1...a_p}(r,g)=T^1_{a_1b_1}...T^1_{a_p b_p}M^{b_1...b_p}\left(rg^{-1}\mathbf{e_3} \right).
\ee
Then, remembering that the dependence of $T^l_{ab}$ on the third Euler angle is of the form $e^{ia\psi}$ and noting that $M^{b_1...b_p}\left(rg^{-1}\mathbf{e_3} \right)$ is independent of $\psi$, we find that $\mathcal{M}_{a_1...a_p}(r,g)$ is proportional to $e^{iA\phi}$, with $A=a_1+...+a_p$. 

We now restrict to $S^2$ by setting $
r=R_0$,
with $R_0$ a constant, and setting all tensor components in the $\mathbf{\hat{r}}$ direction to vanish. We will take the index $a$ to run over $-1,+1$ for the remainder of the paper. In this basis, the metric on the sphere $\sigma$ is given by 
\be
\sigma_{ab}=\begin{pmatrix}0&-1\\-1&0\end{pmatrix}.
\ee

Finally, we can expand into generalised spherical harmonics
\be
\mathcal{M}_{a_1...a_p}(g)=\sum_{l\geq A,|m|\leq l} \left(\mathcal{M}_{a_1...a_p}\right)_l^m T^l_{Am}(g),
\ee
with $A=a_1+...+a_p$, as explained below equation \eqref{sum_expl} and $T^l_{ab}$ introduced in \eqref{gen_sph_harm}. We also have:
\be
\left(\mathcal{M}_{a_1...a_p}\right)_l^m =\tfrac{2l+1}{8\pi^2}\int_{g\in\text{SO(3)}} \mathcal{M}_{a_1...a_p}(g) \bar{T}^l_{Am}dg.
\ee 
The usual orthogonality relations for the Wigner matrices induce similar relations for the generalised spherical harmonics:
\be
\label{ortho_gen_harm}
\int_{g\in \text{SO(3)}} \bar{T}^l_{ab}(g)T^{l'}_{a'b'}(g) \ dg=\tfrac{8\pi^2}{2l+1}\delta_{ll'}\delta_{aa'}\delta_{bb'}.
\ee
In terms of the Euler angles, we can write the above as
\be
\int \bar{T}^l_{ab}(\phi,\theta,\psi)T^{l'}_{a'b'}(\phi,\theta,\psi)\sin\theta d\theta d\phi=\tfrac{8\pi^2}{2l+1}\delta_{ll'}\delta_{aa'}\delta_{bb'}.
\ee 
We are now ready to introduce the tensor convolution, which can be seen as the generalisation of the operator  defined in \eqref{DH_convo} for scalars:
\be 
\label{tensor_convo_def}
\begin{aligned}
k_{a_1...a_m}\cdot f_{b_1...b_n}(\omega)=&\left(\int dg\ k_{a_1...a_m}(g)\Lambda^A(g)\right)f_{b_1...b_n}(\omega)\\
=&\int dg\ k_{a_1...a_m}(g)[X^Af]_{b_1...b_n}\left(g^{-1}\omega \right).
\end{aligned}
\ee
Here,  $\omega\in\text{SO(3)}$  and $\Lambda^A(g)$ is an operator induced by the action of SO(3) on the sphere,  weighted by $A=a_1+\cdots+a_m$.  The operator $X^A$ is defined through its action on the generalised spherical harmonics, $[X^Af]_{b_1...b_n}=(f_{b_1...b_n}){}^{m}_{l}[X^AT]^l_{Bm}\left(\omega \right)$, where
\be 
\label{X_definition}
[X^AT]^l_{Bm}\left(\omega \right):={\Omega}_{(A,B)}^l\  T^l_{B+A,m+A}(\omega).
\ee 
Here the prefactor
\be
\Omega_{(A,B)}^l=\tfrac{\Omega_{-AB}^l}{\Omega_0^l}, 
\ee
with 
\be
\label{omega_def}
\Omega_N^l=\sqrt{\tfrac{(l+N)(l-N+1)}{2}},
\ee
is introduced for convenience, as it will allow for the correct matching of symmetries between  Yang-Mills  and the gravity theory, as shown in \autoref{dcsphere}. It has the property that $\Omega_{N}=\Omega_{-N+1}$.

We now wish to check the factorisation property of the convolution. We Fourier transform
\be
\label{fact_proof_1}
\begin{aligned}
\left(k_{a_1...a_m}\cdot f_{b_1...b_n}\right)_l^m=&\tfrac{2l+1}{8\pi^2}\int_{\omega\in\text{SO(3)}}
k_{a_1...a_m}\cdot f_{b_1...b_n}(\omega)\bar{T}^l_{A+B,m}(\omega)d\omega\\
=&\tfrac{2l+1}{8\pi^2}\int_{\omega,g\in\text{SO(3)}}  k_{a_1...a_m}(g)[X^Af]_{b_1...b_n}\left(g^{-1}\omega \right)\bar{T}^l_{A+B,m}(\omega) dgd\omega,
\end{aligned}
\ee  
with $B=b_1+...+b_n$. We now expand each factor into generalised spherical harmonics to get
\be
\label{fact_exp_k}
 k_{a_1...a_m}(g)=\sum_{l',m'} \left( k_{a_1...a_m} \right)_{l'}^{m'} T^{l'}_{Am'}(g)
\ee
and 
\be 
\begin{aligned}
\,[X^Af]_{b_1...b_n}\left(g^{-1}\omega \right)=&\sum_{l'',m''}\left(f_{b_1...b_n}\right)_{l''}^{m''}[X^AT]^{l''}_{Bm''}\left(g^{-1}\omega \right)\\
=&\sum_{l'',m''}\Omega^{l''}_{(A,B)}\left(f_{b_1...b_n}\right)_{l''}^{m''}
T^{l''}_{B+A,m''+A}(g^{-1}\omega),
\end{aligned}
\ee  
where to get to the second line we made use of the definition \eqref{X_definition}. Next, using the relation between the generalised spherical harmonics and the Wigner matrices \eqref{gen_sph_harm}, together with the factorisation property of representations, one can rewrite the above as:
\be
\label{fact_exp_f}
\,[X^Af]_{b_1...b_n}\left(g^{-1}\omega \right)=\sum_{l'',m''}\Omega^{l''}_{(A,B)}\left(f_{b_1...b_n}\right)_{l''}^{m''}\sum_{\tilde{m}}T^{l''}_{A+B,\tilde{m}}(\omega)\bar{T}^{l''}_{m''+A,\tilde{m}}(g).
\ee
Finally, we can plug \eqref{fact_exp_k} and \eqref{fact_exp_f} back into \eqref{fact_proof_1}, and use the orthogonality relations of the generalised spherical harmonics \eqref{ortho_gen_harm} to get
\be 
\label{tensor_convo_fact}
\left(k_{a_1...a_m}\cdot f_{b_1...b_n}\right)_l^m=\tfrac{8\pi^2}{2l+1}\Omega^l_{(A,B)}
\left( k_{a_1...a_m} \right)_l^m \left(f_{b_1...b_n}\right)_l^0,
\ee
thus proving the factorisation property of the spherical convolution for tensors. As in the scalar case, we notice the asymmetry due to the existence of distinct left/right convolutions.

From herein we will use this convolution in the definition of the $\circ$-product in \eqref{def}. It can be shown that when this convolution is used  in \eqref{def} the isometry transformations on the factors induce an isometry transformation on the product. Crucially the pseudo-inverse spectator scalar is needed here, just as it was in the flat space case.

\section{Gravity $=$ Gauge $\times$ Gauge on $S^2$}\label{dcsphere}
We now proceed to setting up a dictionary which, taking as imput the symmetries of the  Yang-Mills  BRST system:
\be 
Q A_a=\nabla_a c,\quad Qc=0,\quad Q\bar{c}=\tfrac{1}{\xi}G(A)
\ee 
correctly reproduces those of the gravitational BRST system:
\be 
\label{sphere_grav_BRST}
\begin{array}{llllllllllll}
Qh_{ab}&=&2\nabla_{(a}c_{b)},\quad &Qc_a&=&0, \quad &Q\bar{c}_a&=&\tfrac{1}{\xi^{(h)}}G_a[h,\varphi],\quad\  \\ [5pt]
QB_{ab}&=&2\nabla_{[a}d_{b]},\quad &Qd_a&=&\nabla_a d, \quad & Q\bar{d}_a&=&\tfrac{1}{\xi^{(B)}}G_a[B,\eta], \quad\ \\[5pt]
Q\varphi&=&0.\\
\end{array}
\ee
A key result is that when working in the basis defined in \eqref{Burridge_basis} and \eqref{go_to_helicity_basis}, we have
\be 
\left(\nabla_a \mathcal{M}_{a_1...a_n} \right)_l^m =\frac{1}{R_0}\Omega_{A+\tfrac{a+1}{2}}^l
\left(\mathcal{M}_{a_1...a_n} \right)_l^m,
\ee
with $A=a_1+...a_n$ and $\Omega_N^l$ defined in \eqref{omega_def}. Using the above in conjunction with the definition of the convolution \eqref{tensor_convo_def}  and $\circ$-product  \eqref{def} we get: 
\be \label{divrule}
\begin{aligned}
V_a\circ\nabla_b s=&\nabla_b\left(V_a\circ s\right)=\left(\nabla_b V_a\right)\circ s, \\
\nabla_a s\circ V_b=&\nabla_a\left(s\circ V_b \right)=s\circ\left(\nabla_a V_b\right).
\end{aligned}
\ee
Note that the prefactor $\Omega^l_{(A,B)}$ appearing in \eqref{tensor_convo_fact} is essential for obtaining the above derivative rule.
Then one can see that the ans\"atze
\be
\label{sphere_simple_dict}
\begin{aligned}
h_{ab}=&A_a\circ\tilde{A}_b+A_b\circ\tilde{A}_a+\gamma\sigma_{ab}\psi,\\
B_{ab}=&A_a\circ\tilde{A}_b-A_b\circ\tilde{A}_a,
\end{aligned} 
\ee
with $\gamma$ an arbitrary constant and $Q\psi=0$, where $\psi$ is for the moment left undetermined, correctly reproduce the graviton and two-form symmetries, and allow us to read off their respective ghost dictionaries:
\be
\begin{aligned}
c_a=&c\circ\tilde{A_a}+A_a\circ\tilde{c},\\
d_a=&c\circ\tilde{A_a}-A_a\circ\tilde{c}.
\end{aligned} 
\ee
The dictionaries for the invariant combination $\psi$ and the dilaton $\varphi$ are dependent on the choice of gauge fixing functional. For instance, if we pick
\be\label{difficult} 
G[A]\equiv\nabla^a A_a,\quad G[\tilde{A}]\equiv\nabla^a \tilde{A}_a,  
\ee
then the simplest dictionaries for $\psi$ and $\varphi$ are
\be
\label{sphere_simple_dilaton_1}
\psi=\varphi = A^a\circ\tilde{A}_a + \xi c^\alpha\circ\tilde{c}_\alpha,
\ee
with the OSp(2) ghost singlet $c^\alpha\circ\tilde{c}_\alpha$ defined in \eqref{ghost_singlet}. Note, \eqref{sphere_simple_dilaton_1} is the unique solution in the absence of non-local derivative $\Box^{-1}$ terms.  The gravity gauge-fixing condition implied by \eqref{difficult} is subtle\footnote{Specifically, the graviton gauge-fixing functional is 
\be \label{sphereG}
G_\pm[h,\varphi]=\square_{\mp}\nabla^b\tfrac{1}{\square_{\mp}}h_{b\pm}-\tfrac{1}{2}\nabla_\pm h+
\left[\left(1+\gamma\right)\nabla_\pm- \gamma\square_{\mp}\nabla_\pm\tfrac{1}{\square_{\mp}} \right]\varphi,
\ee
where we introduced $\square_{\pm}=P^{ab}_{\pm}\nabla_a\nabla_b$ with $P_{\pm}^{ab}=\tfrac{1}{2}\left(\sigma^{ab}\pm i\varepsilon^{ab}\right)$. We note that for a flat background metric  one recovers \eqref{flat_grav_gauge_simple} from \eqref{sphereG}. One could also imagine recovering \eqref{flat_grav_gauge_simple} by taking an  $R_0\rightarrow \infty$ limit of a punctured sphere for functions that are sufficiently damped  as they approach the puncture.}, principally because the derivative rule given in \eqref{divrule} does not hold for higher rank tensor due to the non-trivial connection.

A different gauge choice, which  maps to a simple de Donder-like gauge choice on the gravity side, is
\be 
\label{ugly_gf}
\begin{aligned}
G'[A]\equiv&\left(-\nabla^a + \tfrac{2}{\square}\nabla^a\square\right) A_a=\left(1+\tfrac{2}{R^2_0\square} \right)\nabla^a A_a,\\
 G'[\tilde{A}]\equiv&\left(-\nabla^a  +  \tfrac{2}{\square}\nabla^a\square\right) \tilde{A}_a=
\left(1+\tfrac{2}{R^2_0\square} \right)\nabla^a \tilde{A}_a.
\end{aligned}
\ee
Then
\be
\label{sphere_simple_dilaton_2}
\psi'=\varphi'=-A^a\circ\tilde{A}_a + \xi  c^\alpha\circ\tilde{c}_\alpha
+\tfrac{2}{\square}\left(\square A^a \right)\circ\tilde{A}_a.
\ee
 Note that for a flat background the two gauge choices coincide and they are identical to the gauge fixing choice in \eqref{flat_gauge_choice}. In this case, the dictionaries \eqref{sphere_simple_dict}, \eqref{sphere_simple_dilaton_1} and \eqref{sphere_simple_dilaton_2} also reduce to the ones in \eqref{flat_simple_dict}, with the spherical convolution replaced by the standard flat-space convolution.

Finally, it is easy to check that
\be
Q c_a =0
\ee
as required, while the transformation of $d_a$ gives, using \eqref{sphere_grav_BRST}
\be
d=-2 c\circ\tilde{c}.
\ee 
We are now ready to set up the gauge-fixing map. We first read off the anti-ghost dictionaries from the ghost ones:
\be 
\begin{aligned}
\bar{c}_a=&\bar{c}\circ\tilde{A_a}+A_a\circ\tilde{\bar{c}},\\
\bar{d}_a=&\bar{c}\circ\tilde{A_a}-A_a\circ\tilde{\bar{c}}.
\end{aligned} 
\ee
Then, making use of the relation between their BRST transformations and the gravitational gauge fixing functional \eqref{sphere_grav_BRST}, and setting $\xi^{(h)}=\xi^{(B)}=\xi$, we read off the gravitational gauge fixing functional associated with \eqref{ugly_gf},
\be 
\begin{aligned}
G_a[h,\varphi]=&\nabla^b h_{ba}-\tfrac{1}{2}\nabla_a h+ \nabla_a \varphi,\\
G_a[B,\eta]=&\left( -\nabla^b+2\square \nabla^b\tfrac{1}{\square}\right)B_{ba}-\nabla_a\eta,
\end{aligned}
\ee
where $\eta$ is identified with $\xi\left(c\circ\tilde{\bar{c}}+\bar{c}\circ\tilde{\bar{c}}\right)$ and we imposed $\xi=\xi^{(h)}=\xi^{(B)}$, as before. Again, note that for  flat background  we recover the gauge mapping presented in \eqref{flat_grav_gauge_simple}, with $D=2$.

\section{The $D=3$ Dimensional Einstein-Static Universe}\label{static}
We can extend our set-up to work on the mostly plus signature Lorentzian $\mathds{R}\times S^2$, with $\mathds{R}$ corresponding to the time dimension. The background solution is the Einstein-Static universe, a simple GR solution in $D=3$ dimensions which exhibits spherical symmetry. Perturbations around this background will not possess full 3D covariance. All our fields are now functions of time, so in addition to the spherical convolution, one needs to perform a standard convolution over the flat time dimension,
\be 
[f\cdot g](t_0)=\int f(t_0-t)g(t) dt . 
\ee
In this section, we take $\circ$ to denote this particular 3D convolution in conjunction with \eqref{def} (note that the bi-adjoint spectator scalar is now also a function of time). 
The time component of the gauge field, $A_t$, transforms as
\be 
Q A_t=\partial_t c
\ee  
under BRST and is a scalar from the perspective of the 2-sphere. The $D=1+2$ graviton ansatz is given by  
\be 
\begin{aligned}
h_{ab}=&A_a\circ\tilde{A}_b+A_b\circ\tilde{A}_a+\gamma\sigma_{ab}\psi_{(3)},\\
h_{ta}=&A_t\circ\tilde{A}_a+A_a\circ\tilde{A}_t,\\
h_{tt}=&2A_t\circ\tilde{A}_t-\gamma\psi_{(3)},
\end{aligned}
\ee
which are a vector and a scalar, respectively, from the perspective of $S^2$. They transform as expected under BRST:
\be 
\begin{aligned}
Q h_{ta}=&\partial_t c_a+\partial_a c_t,\\
Q h_{tt}=&2\partial_t A_t,
\end{aligned}
\ee
which allows us to read off the $t$ component of the graviton ghost,
\be
c_t=A_t\circ \tilde{c}+c\circ\tilde{A}_t.  
\ee
This satisfies $Q c_t=0$, as needed. For the two-form we will similarly have the additional $S^2$ vector:
\be
B_{ta}=A_t\circ\tilde{A}_a-A_a\circ\tilde{A}_t, 
\ee  
which transforms as
\be 
Q B_{ta} =\partial_t d_a-\partial_a d_t,
\ee
thus allowing us to read off
\be
d_t= c\circ \tilde{A}_t - A_t\circ\tilde{c}
\ee
and we have $Q d_t=\partial_t d$ as needed. Finally, we can modify the gauge-fixing term in \eqref{ugly_gf} by a term depending on $A_t$
\be 
\label{ugly_gf_with_t}
\begin{aligned}
G'[A]\equiv&- \partial_t A_t+\left(-\nabla^a+ \tfrac{2}{\square}\nabla^a\square\right) A_a= - \partial_t A_t +  \left(1+\tfrac{2}{R^2\square} \right)\nabla^a A_a,\\
 G'[\tilde{A}]\equiv&- \partial_t \tilde{A}_t+\left(-\nabla^a +  \tfrac{2}{\square}\nabla^a\square\right) \tilde{A}_a= - \partial_t \tilde{A}_t+
\left(1+\tfrac{2}{R^2\square} \right)\nabla^a \tilde{A}_a.
\end{aligned}
\ee
 The dilaton dictionary is correspondingly  modified to 
\be 
\varphi_{(3)}=\psi_{(3)}= - A_t\circ\tilde{A}_t -A^a\circ\tilde{A}_a 
+\tfrac{2}{\square}\left(\square A^a \right)\circ\tilde{A}_a+ \xi c^\alpha\circ\tilde{c}_\alpha.
\ee 
 The gravitational gauge fixing functionals associated with \eqref{ugly_gf_with_t} are then found to be:
\be 
\begin{aligned}
G_a[h,\varphi]=&-\partial_t h_{ta}+ \nabla^b h_{ba}-\tfrac{1}{2}\nabla_a h+ \left(1+\tfrac{\gamma}{2} \right)\nabla_a\varphi_{(3)},\\
G_t[h,\varphi]=&-\tfrac{\nabla^a\square\nabla_a}{\square^2}\partial_t h_{tt}+\left(-\nabla^a+\tfrac{2}{\square}\nabla^a\square\right)h_{at}\\
&-\left(\tfrac{2\nabla^a\square\nabla_a}{\square^2} -1\right)\tfrac{1}{2}\partial_t h+\left[1+\tfrac{\gamma}{2}\left(\tfrac{4\nabla^a\square\nabla_a}{\square^2} -3\right) \right]\partial_t\varphi_{(3)},  \\
G_a[B,\eta]=&-\partial_t B_{ta}+\left(-\nabla^b+2\square \nabla^b\tfrac{1}{\square}\right)B_{ba}-\nabla_a\eta,\\
G_t[B,\eta]=&\left( -\nabla^a+\tfrac{2}{\square}\nabla^a\square\right)B_{at}-\partial_t \eta. 
\end{aligned}
\ee
Just as in the case of $S^2$, we see that on a flat background takes us to the results in \autoref{rev}.

\section{Conclusions}

By introducing a convolution of tensor fields on $S^2$ we have shown that the ``gauge $\times$ gauge'' product of \cite{Anastasiou:2014qba} can be applied to fields on $S^2$. The convolution was defined to satisfy two key properties with respect to this goal: (i) it is multiplicative in Fourier space; (ii) it commutes with the covariant derivative for convolutions with a (suitably well-behaved) scalar field, reflecting the familiar flat-space derivative rule \eqref{divrule}. However, it cannot, and indeed should not, commute  for higher rank tensors. 

Including the BRST complex in the ``gauge $\times$ gauge''  product, we find that the BRST transformations and gauge-fixing conditions of the gravity theory are derived from those of the Yang-Mills factors to linear order. The dictionary relating the gravity fields to the left/right Yang-Mills fields  is unique if one restricts to only local derivative operators.  On the other hand, the gauge-fixing functions are seen to typically involve non-local derivative operators on either the gauge or gravity side. Nonetheless, they reduce to the expected  conditions for a flat background. Finally, the product was shown to almost trivially extend to the $D=1+2$ Einstein-static universe. 

There are a number of possible further directions. Ultimately, a definition of the convolution robust enough for (reasonable) bundles over arbitrary differentiable manifolds would be desirable. Here, as a first step, we considered the simplest coset space $S^2\cong \SO(3)/\SO(2)$, where the fact that we have a transitive group action on $S^2$ facilitated the definition of the convolution. With  this in mind, the next class to consider are the group manifolds. In this case one has a \emph{free} transitive group action on the manifold, which allows for an immediate generalisation of the familiar flat-space, which is nothing but the very simplest example of  a group manifold,  convolution. In this sense, the group manifolds are actually more straightforward than the $S^2$ convolution treated here. The first  non-trivial example is given by $S^3\cong \SO(3)$ (with a marked point). Since the group elements correspond to points, one can  directly use the Wigner $D$-matrices to establish the required properties. This, moreover, would allow us to move up in dimension  to the physically relevant case  of $D=1+3$, for the Einstein-static universe. One could also consider spinor fields using the double-cover $\text{SU}(2)$, allowing for the product of super\footnote{The inclusion of adjoint fermions in the gauge theories implies supersymmetry for both  BCJ duality \cite{Chiodaroli:2013upa, Weinzierl:2014ava} and the field ``gauge $\times$ gauge'' product \cite{Anastasiou:2014qba}.} Yang-Mills multiplets on $S^3$. In this case, in addition to the linearised diffeomorphism ghost  arising from the product of  gauge potentials and ghosts, we would expect a commuting  local supersymmetry ghost arising from the product of the spinor and ghost fields.


\acknowledgments

The Authors  gratefully acknowledge illuminating discussions with Michael J.~Duff, Andres Luna,  Ricardo Monteiro, Denjoe O'Connor, Donal O'Connell and  Chris White. The work of LB is supported by a Sch\"odinger Fellowship. The work of IJ is supported by an IRC Postdoctoral Fellowship. The work of SN is supported by a Leverhulme Research Project Grant.



\providecommand{\href}[2]{#2}\begingroup\raggedright\endgroup

\end{document}